\documentclass[journal]{IEEEtran}
\usepackage{amsmath,amsfonts}
\usepackage{algorithmic}
\usepackage{algorithm}
\usepackage{array}
\usepackage[caption=false,font=footnotesize,labelfont=rm,textfont=rm]{subfig}
\usepackage{textcomp}
\usepackage{stfloats}
\usepackage{url}
\usepackage{verbatim}
\usepackage{graphicx}
\usepackage{cite}
\usepackage{color}
\usepackage{booktabs}
\usepackage{multirow}
\usepackage{makecell}
\usepackage{multicol}
\usepackage[switch]{lineno}
\begin{document}
\title{A One-dimensional HEVC video steganalysis method using the Optimality of Predicted Motion Vectors}

\author{Jun Li, Minqing Zhang, Ke Niu, Yingnan Zhang, Xiaoyuan Yang.
        % <-this % stops a space
\thanks{This work was supported by the National Natural Science Foundation of China (Grant No.62272478, No.62202496, No.62102450). (Corresponding author: Minqing Zhang.)

The executable program related to this paper can be downloaded  from:https://github.com/lijun9250lj/HEVC-MVPO-Video-steganalysis.
	
The authors are with the College of Cryptography Engineering in Engineering University of the Chinese People’s Armed Police Force, Xi’an, 710086, China; (e-mail: lijun9250lj@163.com, api\_zmq@126.com, niuke@163.com, zyn583@163.com, yxyangyxyang@163.com).}% <-this % stops a space
%\thanks{Manuscript received xxx}
}

\markboth{ }%
{Shell \MakeLowercase{\textit{Jun Li et al.}}: Motion Vector-Domain Video Steganalysis Exploiting Skipped Macroblocks}

%\IEEEpubid{0000--0000/00\$00.00~\copyright~2022 IEEE}

\maketitle

%页码
%\pagestyle{empty}
\pagestyle{plain}
\thispagestyle{plain}

\begin{abstract}
Among steganalysis techniques, detection against motion vector (MV) domain-based video steganography in High Efficiency Video Coding (HEVC) standard remains a hot and challenging issue. For the purpose of improving the detection performance, this paper proposes a steganalysis feature based on the optimality of predicted MVs with a dimension of one. Firstly, we point out that the motion vector prediction (MVP) of the prediction unit (PU) encoded using the Advanced Motion Vector Prediction (AMVP) technique satisfies the local optimality in the cover video. Secondly, we analyze that in HEVC video, message embedding either using MVP index or motion vector differences (MVD) may destroy the above optimality of MVP. And then, we define the optimal rate of MVP in HEVC video as a steganalysis feature. Finally, we conduct steganalysis detection experiments on two general datasets for three popular steganography methods and compare the performance with four state-of-the-art steganalysis methods. The experimental results show that the proposed optimal rate of MVP for all cover videos is 100\%, while the optimal rate of MVP for all stego videos is less than 100\%. Therefore, the proposed steganography scheme can accurately distinguish between cover videos and stego videos, and it is efficiently applied to practical scenarios with no model training and low computational complexity.
\end{abstract}


\section{Introduction}
Steganography aims to embed secret messages in multimedia such as picture, audio, and video without arousing suspicion, thus enabling covert communication. On the other hand, the purpose of its adversary steganalysis is to detect the presence of embedded secret messages in ordinary media. Video is the ideal cover for steganography, and there are different steganography methods according to the embedding location\cite{ref1} in video, mainly intra-frame prediction modes\cite{ref2,ref3,ref4}, inter-frame prediction modes\cite{ref5,ref6,ref7,ref8}, MVs\cite{ref9,ref10,ref_Liu_tcsvt_2022}, transformation coefficients\cite{ref11,ref12}, etc. Since many MVs are available for message embedding in video coding, more methods are based on the MV domain. Thus MV-based steganalysis technique is a current research hotspot.

With the gradual popularization and application of the HEVC standard\cite{ref13}, the research of MV-based video steganography and steganalysis techniques based on the HEVC is particularly important. Yang et al.\cite{ref14} proposed a steganography method based on MV space coding for HEVC. They gave the construction and coding process of MV space. They defined the mapping relationship between the set of MVs and the points in the space, which can achieve the effect of embedding a 2N+1 binary number by changing at most one component among N MV components. Guo et al.\cite{ref10} first counted the motion trend of each frame and established a Motion Trend Based (MTB) mapping strategy between the MV and the binary bitstream, and then used the Sum of Transform Difference (SATD) difference before and after the MV modification as steganographic distortion for message embedding. Hu et al.\cite{ref16} first proposed a new steganography method Steganography by Advanced Motion Vector Prediction (SAMVP) using the AMVP technique in HEVC. SAMVP uses the MVP index in the AMVP technique of inter-frame prediction as the embedding cover, which has a sizeable embedding capacity and is lossless. Liu et al.\cite{ref17} proposed the Adaptive-SAMVP (A-SAMVP) based on SAMVP by defining the cost function and combining it with Syndrome Trellis Code (STC)\cite{ref18}. Since AMVP encodes MVs by MVP index values and MVDs, A-SAMVP embeds the information in the index values of the candidate list and uses the code rate difference between two candidate MVPs to define the cost function. The overall performance of the algorithm is improved.

The MV-based steganography algorithm is a modification of the MV and its associated information, which inevitably destroys the optimality of specific parameters in the video coding process, so some traditional H.264/AVC-based steganalysis methods are still effective to some extent in HEVC, such as Adding or Subtracting One(AoSO)\cite{ref19}, Near Perfect Estimation for Local Optimality (NPELO)\cite{ref20}, Motion Vector Consistency (MVC)\cite{ref21}. Nevertheless, to improve the detection efficiency of steganographic algorithms for HEVC, researchers have tried to design steganalysis features by combining the characteristics of HEVC. Shanableh et al.\cite{ref22} extended the idea of the MVC approach to HEVC. They redefined the concept of block group based on the coding depth according to the characteristics of HEVC standard and proposed the feature sets based on MV non-consistency. Huang et al.\cite{ref23} introduced the convolutional neural network to the MV domain video steganalysis based on the HEVC standard and proposed the Video Steganalysis Residual Network (VSRNet) structure. The method constructs independent VSRNet sub-networks for different embedding rates and finally connects all sub-network structures to form a quantitative steganalysis convolutional neural network. Based on VSRNet, they further introduce information such as Selection-Channel-Aware\cite{ref24} and MVD\cite{ref25} to improve the performance of steganalysis. In the new type of MV modification strategy\cite{ref16,ref17} based on the HEVC standard, it is possible to modify only the MVP index without changing the MV itself. So, the traditional MV-based steganalysis features are ineffective for this new type of steganography algorithm. However, if the MVP index is modified, the local optimality of the MVP in the candidate list may be destroyed. Based on this observation, Liu et al.\cite{ref26} constructed steganalysis features based on local optimality on the MVP candidate list and MV, and they proposed the Local Optimality in Candidate List (LOCL) method, which effectively improves the detection performance in HEVC. 

However, existing MV-based video steganalysis methods still have some significant shortcomings. Firstly, the current methods ignore the disturbance caused by MV steganography to the local optimality of MVP in HEVC, which leads to low detection effectiveness in current video steganalysis. Secondly, existing methods are based on machine learning models that require a significant amount of training to achieve an ideal detection model. However, these trained steganalysis models often have low robustness, as they tend to exhibit noticeable performance degradation in the presence of cover or algorithm mismatches.

Based on the above analysis, this paper focuses on the local optimality of the MVP candidate list in HEVC and fully explores the statistical differences before and after message embedding to design the steganalysis feature. First, either the traditional steganography of modifying MVDs or the new steganography of modifying MVP indexes may have perturbations on the local optimality of MVP. Second, we propose a steganalysis feature with a dimension of only one based on the local optimality of the MVP, which is defined as the optimality rate of the MVP in HEVC codestreams. The optimality rate of the MVP is 100\% in all cover videos and below 100\% in all stego videos. Based on this feature, we can accurately determine whether or not the video has been modified by steganography.

The main contributions of this paper can be summarized as follows:
\begin{enumerate}
	\item{We analyzed whether information embedding is based on MVP index or MVD, it may cause disturbance to the optimality of MVP in AMVP technique.}
	\item{The optimal rate of the MVP with a dimension one is defined as the steganalysis feature, which is the lowest dimension among the existing MV domain steganalysis features.}
	\item{The proposed scheme does not require redundant model training during execution, so our method has the advantages of low computational complexity and applicability to practical application scenarios.}
\end{enumerate}

The rest of the paper are organized as follows. The second part introduces the basics knowledge of AMVP technology. The third part analyzes the effect of on the MVP by message embedding based on MVP index and MVD, and defines the optimal rate of MVP as a feature for steganalysis. Then it is proved theoretically that the optimal rate is 100\% in the cover video and below 100\% in the stego video. The experimental results and analysis are given in the fourth part. Finally, the paper is concluded.

\section{Preliminaries}
\subsection{The Technology of Advanced Motion Vector Prediction}
AMVP is an MV prediction technique for inter-frame encoding proposed in HEVC. AMVP uses the correlation of MVs in the spatial and temporal domains to build a list of candidate MVPs (including $ mvp_{0} $ and $ mvp_{1} $) for the current coding Prediction Unit (PU). The optimal MVP $ mvp_{idx},idx\in\{0,1\} $ is selected from the candidate list, and the final optimal MV $ mv $ is obtained by whole-pixel and sub-pixel motion estimation starting from the $ mvp $. Then The MVD $ mvd $ is obtained by 
\begin{equation}
	\label{formula1}
	mvd=mv-mvp_{idx}. 
\end{equation}
$ mvd $ is finally encoded using 0-th order Exp-Golomb codes\cite{ref13}. The decoder recovers the $ mv $ of the current PU by building the same list of candidate MVPs, and only needs the index value $ idx $ of $ mvp $ in the candidate list and the $ mvd $ . So that the recovered $ mv=mvd+mvp_{idx} $. 

\subsection{The Local Optimality of the MVP }
HEVC adopts the Lagrangian optimization algorithm to achieve encoding control in selecting the optimal MVP from the candidate list. The definition of Lagrangian rate-distortion is as follows:
\begin{equation}
	\label{formula2}
	{J_{motion}}(mv) = D + \lambda *R,
\end{equation}
where $ D $ represents the pixel distortion caused by encoding using the current $ mv $. The distortion $ D $ is usually calculated using the Sum of Absolute Difference (SAD) or Hadamard Sum of Absolute Transformed Difference (SATD). $\lambda$ is a Lagrangian parameter that controls the balance between bit rate and distortion. $ R $ represents the number of bits required to encode the current $ mv $, which is actually the number of bits required to encode $ mvd $ and MVP index $ idx $:
\begin{equation}
	\label{formula3}
	R = Bits(mvd) + Bits(idx) = Bits(mvd) + 1,
\end{equation}
where $ Bits(idx)=1 $ is the number of bits required to encode the $ idx $, and $ Bits(mvd) $ is the number of bits required to encode $ mvd $ using the 0-th order Exp-Golomb codes.

According to the Lagrangian optimization model, without loss of generality, assuming that the optimal MVP selected by the encoder in the candidate list is $ mvp_{idx} $, then $ mvp_{idx} $ must meet \textbf{the local optimality of the MVP}:
\begin{equation}
	\label{formula4}
	{J_{motion}}(mvp_{idx}) \le {J_{motion}}(mvp_{\overline {idx} }),
\end{equation}
where $ \overline {idx} $  represents the values in the set $\{0, 1\} $ that is different from $ idx $. Formula (\ref{formula4}) means the rate-distortion of the MVP corresponding to index $ idx $ must be the smallest in the candidate list. Due to the fact that the optimal MVP has been determined by the encoder during the final confirmation of the MVP, which means the reference block is determined. Therefore, the distortion $ D $ of the two candidate MVPs is the same, so the local optimality of the MVP in Formula (\ref{formula4}) can be simplified as:
\begin{equation}
	\label{formula5}
	R(mv{p_{idx}}) \le R(mv{p_{\overline {idx} }}),
\end{equation}
That is to say the number of bits encoding the optimal MVP $ mvp_{idx} $ is lower than that of another candidate $ mvp_{\overline {idx}} $.

\section{The Proposed Steganalysis Method}
In this section, we first analyze the security risk of the HEVC steganography method using MVP index and MVD, i.e., both of them can perturb the local optimality of the MVP. Then a steganalysis feature is designed based on the optimality of the MVP in AMVP.

\subsection{Motion Vector Domain based Steganography in HEVC}\label{chap:MV-Steganography}
Based on the analysis in the previous section, the Lagrangian rate-distortion optimization model first finds the optimal $ mvp $ from the candidate list of MVPs and then finds the optimal $ mv $ by motion estimation. Thus the selected $ mvp $ is optimal in the sense of rate-distortion in the candidate list. MV-based steganography in HEVC can use the MVP index $ idx $ or MVD $ mvd $ as cover. The effect of these two embedding methods on the optimality of the MVP is analyzed below.

\subsubsection{Using the Index of MVP for Message Embedding}\label{chap:MV-Steganography-idx}

Each PU encoded with the AMVP technique has an MVP index $ idx,idx\in\{0,1\} $. SAMVP \cite{ref16} and A-SAMVP \cite{ref17} are the new type of steganographic approaches with $ idx $ as the cover. In the method of SAMVP, when the secret message is the same as $ idx $, the information of the corresponding PU block does not have to be modified. When the secret message differs from $ idx $, the value of $ idx $ must be modified to $ \overline {idx} $, and then the corresponding $ mvp $ is also modified. According to $ mvd=mv-mvp $, since $ mvp $ is changed while $ mv $ remains unchanged, $ mvd $ correspondingly needs to be modified. According to the analysis of Formula (\ref{formula2}), firstly, as $ mv $ is unchanged, then the corresponding best matching block is not changed, so there is no change in pixel distortion $ D $, i.e., no visual distortion for this message embedding; Secondly, the bit rate has changed, mainly due to the change in the number of bits required for $ mvd $ encoding using 0-th order Exp-Golomb codes. Therefore, although SAMVP is lossless in visual quality, it may increase the bit rate. To reduce the impact of embedding operations on the bit rate, A-SAMVP constructs an adaptive steganographic method using STC. The scheme improves the performance of the steganography method by taking the differences in bit rate before and after message embedding as the cost function.

\begin{figure}[!t]
	\centering
	\subfloat[]{\includegraphics[width=3.5in]{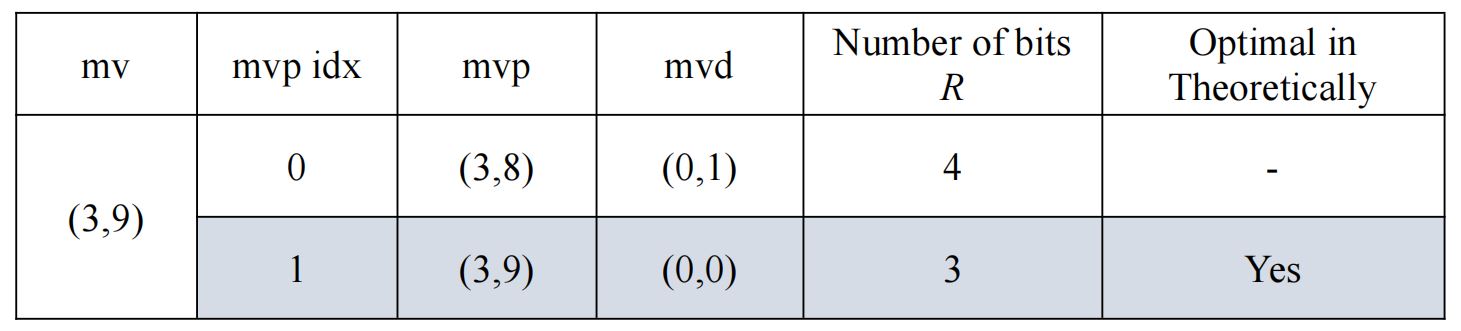}%
		\label{fig1(a)}}
	\hfil
	\subfloat[]{\includegraphics[width=3.5in]{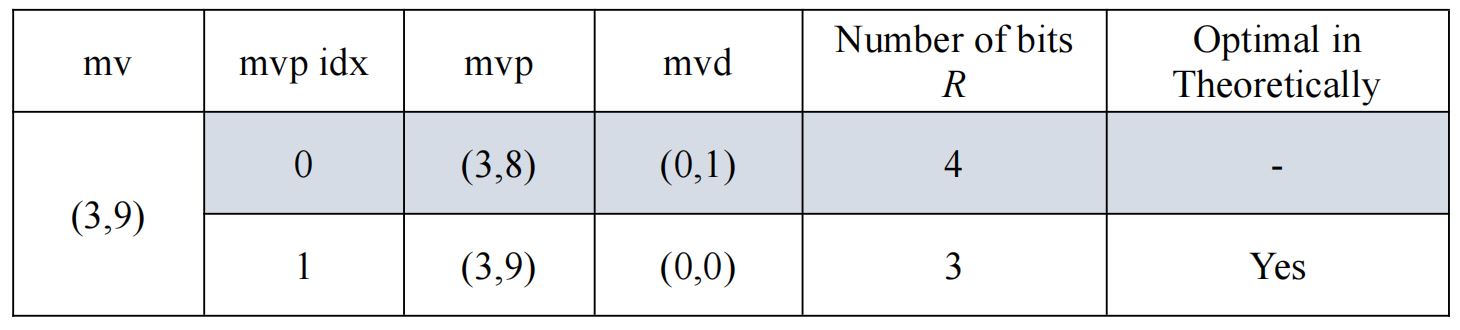}%
		\label{fig1(b)}}
	\hfil
	\subfloat[]{\includegraphics[width=3.5in]{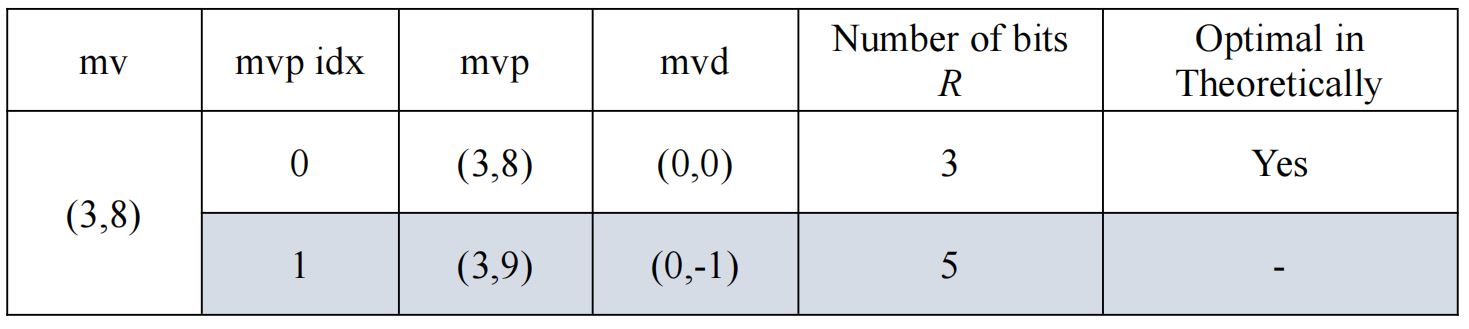}%
		\label{fig1(c)}}
	\caption{An example for the local optimality of MVP in HEVC. The gray background represents the actual situation of MVP observed by the decoding end. (a) Cover video, the MVP satisfies the local optimality. (b) Stego video by modifying the MVP index $ idx $, the MVP do not satisfies the local optimality. (3) Stego video by modifying the $ mvd $, the MVP do not satisfies the local optimality.}
	\label{fig1}
\end{figure}

Although the schemes in SAMVP and A-SAMVP can be lossless in visual quality, an obvious security risk exists. According to the Lagrangian optimization model, the encoder must satisfy the local optimality in Formula (\ref{formula4}) after selecting the optimal $ mvp_{idx} $ from the MVP candidate list. Therefore, if the $ mvp_{idx} $ is artificially modified to $ mvp_{\overline {idx}} $, there will be evident modification traces from the decoder. Fig. \ref{fig1}$ (a), (b) $ show the scenarios where the local optimality of the MVP is corrupted due to message embedding using $ idx $. Fig. \ref{fig1}(a) shows the normal case before message embedding. The optimal $ mv $ by motion estimation is $ (3, 9) $, the two candidates $ mvp_{0} $ and $ mvp_{1} $ in the MVP candidate list are $ (3,8) $ and $ (3,9) $, respectively, and the corresponding $ mvd $ are $ (0,1) $ and $ (0,0) $, respectively. Calculated according to fomula (\ref{formula3}), the number of encoding bits corresponding to the two MVPs is $ 4 $ and $ 3 $, respectively. So the optimal MVP index is $ idx=1 $, i.e., $ mvp_{1} $ will be selected as the optimal MVP. Fig. \ref{fig1}(b) shows the situation after message embedding. Assuming that $ idx $ changes from $ 1 $ to $ 0 $, which means the $ mvp_{0} $ is selected for AMVP. And the number of bits needed to encode the two candidate MVPs at the decoding side is still $ 4 $ and $ 3 $, respectively. As a result, the optimal MVP index $ idx $ should be 1 and $ mvp_{1} $ should be selected as the optimal MVP in theory. However, in practice, the $ idx $ obtained by decoder is $ 0 $, thus destroying the optimality of the MVP.

Based on the above analysis, using the MVP index $ idx $ as the message embedding cover could destroy the MVP's local optimality.

\subsubsection{Using the MVD for Message Embedding}\label{chap:MV-Steganography-mvd}
Using the MVD as a cover for message embedding is a traditional steganography method\cite{ref9,ref14} in H.264/AVC and HEVC. In HEVC, according to $ mvd=mv-mvp $, since $ mvd $ is modified to $ mvd' $ but $ mvp $ remains unchanged, which means the $ mv $ needs to be modified:
\begin{equation}
	\label{formula6}
	mv' = mvd' + mvp,
\end{equation}
That is to say, the optimal matching block corresponding to the current PU has changed. Although these steganography methods do not directly modify the MVP index $ idx $, they may still destroy the MVP's local optimality. This is because the rate-distortion of the corresponding two candidate MVPs will be changed after the $ mvd $ modification. Still, Fig. \ref{fig1}(a) shows the PU of cover video before embedding, and Fig. \ref{fig1}(c) shows the case after the $ mvd $ of this PU block is modified. Suppose the $ mvd $ changes from $ (0, 0) $ to $ (0, -1) $ after message embedding(usually, only one component is modified, and the modification amplitude is one). At this time, $ idx=1 $ remains unchanged, and according to Formula (\ref{formula6}), the corresponding $ mv $ at the decoding side will be changed to $ (3, 8) $. From the decoding end, the $ mvd $ at $ idx=0 $ becomes $ (0, 0) $. According to Formula (\ref{formula3}), the number of bits required to encode $ mvp_{0} $ and $ mvp_{1} $ is $ 3 $ and $ 5 $, respectively. So the optimal MVP index $ idx $ should be $ 0 $ theoretically, but it is actually $ 1 $, thus destroying the optimality of the MVP. From the above analysis, the steganography methods using the MVD as the cover may also destroy the MVP's optimality.

\subsection{The Proposed One-dimensional Steganalysis Feature based on the local optimality of the MVP}\label{chap:Proposed method}
According to the analysis in Section \ref{chap:MV-Steganography}, in the HEVC standard, both the traditional steganography method using MVD as cover and the new steganography method using MVP indexes as cover may perturb the local optimality of the MVP. Based on the observation, this paper defines \textbf{the optimal rate of MVP} as the steganalysis feature:
\begin{equation}
	\label{formula7}
	Optimal(mvp) = \frac{{\sum\limits_{i = 1}^N {\delta ({J_{motion}}(mv{p_{id{x_i}}}),{J_{\min }})} }}{N} \times 100\% ,
\end{equation}
where $ N $ is the total number of all PUs encoded with the AMVP technique in a video sequence. $ \delta $ is a check function, $\delta (x,y){\rm{ = 1}}$ when x equals y and $\delta (x,y){\rm{ = 0}}$ otherwise. $J$ is the Lagrangian rate-distortion calculated according to Formula (\ref{formula2}) , and ${J_{\min }} = \min \{ {J_{motion}}(mv{p_{id{x_i}}}),{J_{motion}}(mv{p_{\overline {id{x_i}} }})\} $.

\textbf{Property 1:} The optimal rate of MVP in the cover video is 100\%.

\textbf{Proof:} According to the HEVC standard, AMVP select the one with the minimum Lagrangian rate-distortion in the MVP candidates list$\{ mv{p_0},mv{p_1}\} $ as the optimal MVP for the current PU. Without loss of generality, assuming that the optimal MVP in PU is $mvp_{0}$, then according to the rate-distortion minimization rule, there must be:
\begin{equation}
	\label{formula8}
{J_{motion}}(mv{p_0}) \le {J_{motion}}(mv{p_1}),
\end{equation}
and then there are:
\begin{equation}
	\label{formula9}
	\begin{array}{l}
		{J_{\min }} = \min \{ {J_{motion}}(mv{p_0}),{J_{motion}}(mv{p_1})\} \\
		\quad \quad  = {J_{motion}}(mv{p_0}),
	\end{array}
\end{equation}
so that $\delta ({J_{motion}}(mv{p_0}),{J_{\min }}) = 1$,and then:
\begin{equation}
	\label{formula10}
	\begin{array}{l}
		Optimal(mvp) = \frac{{\sum\limits_{i = 1}^N {\delta ({J_{motion}}(mv{p_0}),{J_{\min }})} }}{N} \times 100\% \\
		\quad \quad \quad \quad \quad \quad  = \frac{{\sum\limits_{i = 1}^N 1 }}{N} \times 100\%  = 100\% 
	\end{array},
\end{equation}
finally $Optimal(mvp) = 100\% $, and the proof is completed.

\textbf{Property 2:} If the local optimality of MVP of some PUs in the stego video is broken, the optimal rate of the MVP in the stego video is less than 100\%.

\textbf{Proof:} Without loss of generality, it is assumed that AMVP chooses $ mvp_{0} $ as the optimal MVP before message embedding. According to the analysis in Section \ref{chap:MV-Steganography}, message embedding using either the MVP index or the MVD may destroy the local optimality of the MVP.

Case (1). For the steganography method of using the MVP index as cover. If the local optimality of the MVP of some PUs is corrupted after embedding, i.e., the selected optimal MVP in the encoder becomes $ mvp_{1} $ after embedding(see section \ref{chap:MV-Steganography-idx}), and ${J_{motion}}(mv{p_0}) \le {J_{motion}}(mv{p_1})$. So that ${J_{\min }} = \min \{ {J_{motion}}(mv{p_0}),{J_{motion}}(mv{p_1})\}  = {J_{motion}}(mv{p_0})$, and $\delta ({J_{motion}}(mv{p_1}),{J_{\min }}) = 0 < 1$. So the optimal reate of MVP in the Decoder is:
\begin{equation}
	\label{formula11}
		\begin{array}{l}
			Optimal(mvp) = \frac{{\sum\limits_{i = 1}^N {\delta ({J_{motion}}(mv{p_1}),{J_{\min }})} }}{N} \times 100\% \\
			\quad \quad \quad \quad \quad \quad  < \frac{{\sum\limits_{i = 1}^N 1 }}{N} \times 100\%  = 100\% 
		\end{array},
\end{equation}

Case (2). For the steganography method of using the MVD as cover. If the local optimality of the MVP of some PUs is corrupted after embedding, the optimal MVPs selected by these PUs remain unchanged, but according to the analysis in section \ref{chap:MV-Steganography-mvd}, the MVDs have changed. Therefor,  ${J_{motion}}(mv{p_1}) < {J_{motion}}(mv{p_0})$, and ${J_{\min }} = \min \{ {J_{motion}}(mv{p_0}),{J_{motion}}(mv{p_1})\}  = {J_{motion}}(mv{p_1})$. And then there is $\delta ({J_{motion}}(mv{p_0}),{J_{\min }}) = 0$, so the optimal rate of MVP in the Decoder is :
\begin{equation}
	\label{formula12}
	\begin{array}{l}
		Optimal(mvp) = \frac{{\sum\limits_{i = 1}^N {\delta ({J_{motion}}(mv{p_0}),{J_{\min }})} }}{N} \times 100\% \\
		\quad \quad \quad \quad \quad \quad  < \frac{{\sum\limits_{i = 1}^N 1 }}{N} \times 100\%  = 100\% 
	\end{array},
\end{equation}
Combining formulas (\ref{formula11}) and (\ref{formula12}), the proof of Property 2 is completed.

\textbf{Corollary 1:} Given a video sequence, if its optimal rate of MVP is less than 100\%, the sequence is a stego video.

\textbf{Proof:} According to Property 1, if this video is a cover video, its optimal rate of MVP must be equal to 100\%. On the contrary, if its optimal rate of MVP is lower than 100\%, it means that the optimality of the MVP of some PUs is perturbed, which is an abnormal phenomenon. This perturbation comes from the message embedding, so that the video can be judged as stego.

\begin{figure}[!t]
	\centering
	\includegraphics[width=3.5in]{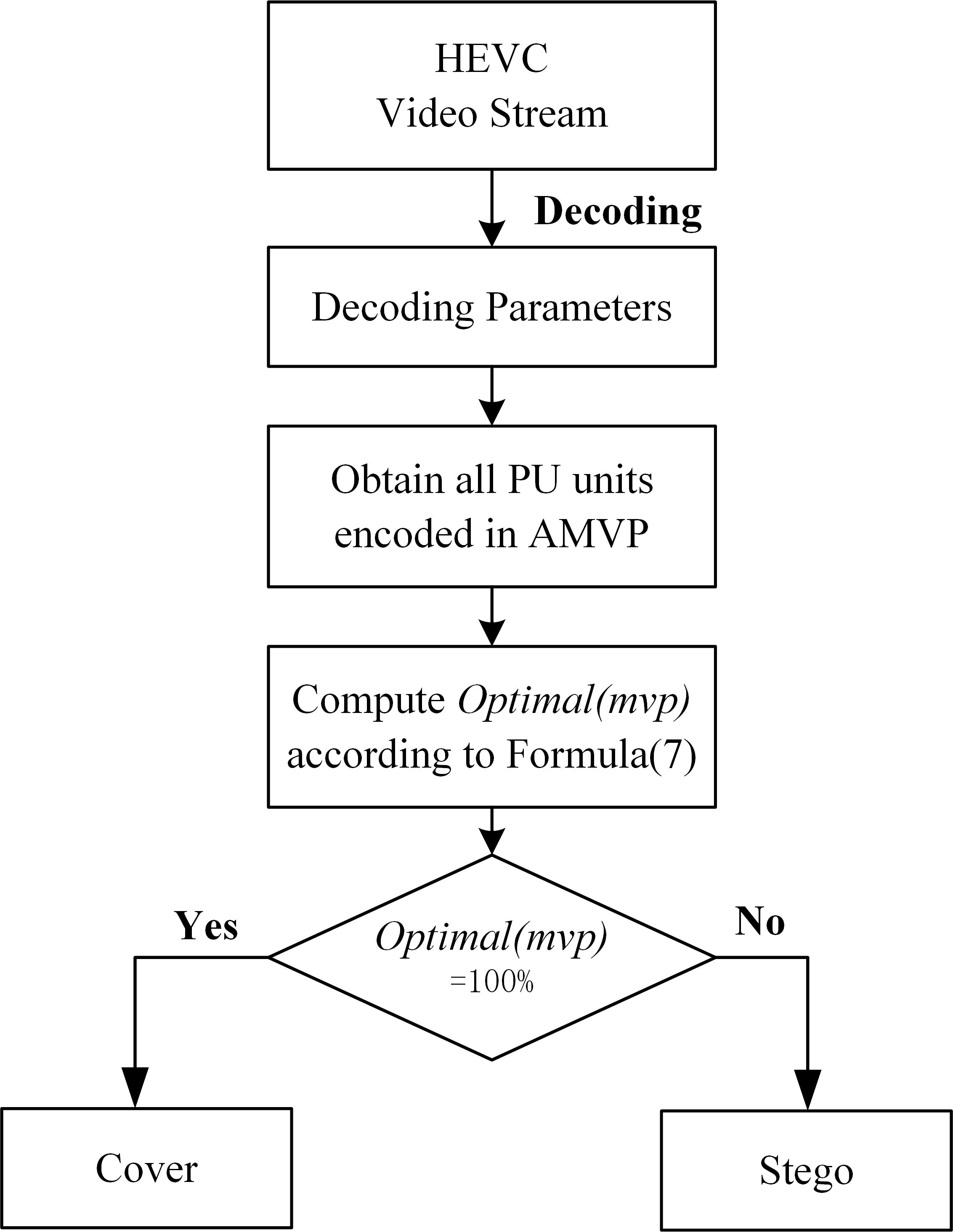}
	\caption{The steganalysis process of the proposed scheme.}
	\label{fig2}
\end{figure}

Through the analysis of Properties 1, 2 and Corollary 1, we can use the optimal rate of MVP $ Optimal(mvp) $ as the steganalysis feature for determining whether the video sequence of HEVC has been modified by steganography. The proposed steganalysis process is shown in Fig. \ref{fig2}. First, a given HEVC compressed video sequence is decoded to obtain the decoding parameters. Next, all PU units encoded using the AMVP technique and their corresponding parameters (MVs, MVP candidate lists, etc.) are collected. Then, the optimal rate of MVP $ Optimal (mvp) $ for the video sequence is calculated based on Formula (\ref{formula7}). Finally, the value of $ Optimal (mvp) $ is used for judgment. If $ Optimal (mvp) = 100\% $, it indicates that the optimal MVPs of all PU units encoded using the AMVP technique are intact, and the video sequence is a normal cover video. If $ Optimal (mvp) < 100\% $, it indicates that the optimal MVPs of some PU units have been damaged, and the video sequence is a stego video.

\section{Experiments and Analysis}
In this section, some different setups are presented to evaluate the performance of the proposed scheme.
\subsection{Experiments Setup}
\subsubsection{Video Databases}
TABLE \ref{tbl_DB} shows two video databases used for the experiments. The database DB1 contains 34 well-known standard test sequences\cite{ref27} with a CIF resolution(352×288), and each video sequence is cut into a fixed length by selecting its first 240 frames (so the total number of frames for experiments is 8160). Another database, DB2, contains 80 standard test sequences with different resolutions(from 416×240 to 2560×1600), which are downloaded from the internet, and each sequence is cut into a fixed length by selecting its first 100 frames (so the number of total frames is 8000). All the video sequences in DB1 and DB2 are stored in uncompressed file format, with YUV 4:2:0 color space.
\begin{table}[t]
	\caption{\label{tbl_DB} Databases with YUV format of test video sequences.}
	\centering
	\renewcommand\arraystretch{1.3}
	\begin{tabular}{ccccc}
		\toprule
		Database  &\makecell{Number of \\Total Frames} & Resolution &\makecell{Number of \\Frames in \\Each Sequence} & \makecell{Number of \\Sequence} \\  
		\toprule 		
		DB1	      &8160	        &352 × 288	  &240   &	34  \\
		\midrule
		\multirow{5}{*}{DB2}   & \multirow{5}{*}{8000}    &2560×1600	&100	&10 \\
		&							&1920×1080	&100	&10 \\
		&							&1280×720	&100	&20 \\
		&							&832×480    &100	&20  \\
		&							&416×240    &100	&20 \\
		\bottomrule                 	
	\end{tabular} 
\end{table}

\subsubsection{Steganography Methods}\label{chap:steganography methods}
Three state-of-the-art typical MV-based steganography methods for HEVC are used for message embedding to evaluate the detectability of video steganalysis in the MV domain. The first one is Yang’s method\cite{ref14}(denoted as Tar1), the second one is Hu’s method\cite{ref16}(denoted as Tar2), and the last one is Liu’s method\cite{ref17}(denoted as Tar3). Due to the different design principles of the above three methods, their embedding capacities are evaluated differently. The embedding strength $ e $ in Tar1 is a decimal whose range is 0 to 1, representing the probability of whether the secret information is embedded in CTU, which shall be set at 0.1, 0.2, 0.3, 0.4, and 0.5. The embedding threshold $ T $ in Tar2 is defined as $T = abs(abs({H_1}) - abs({H_0})) + abs(abs({V_1}) - abs({V_0}))$, where $ H_{0} $ , $ V_{0} $ represents the horizontal and vertical components of $ mvp_{0} $, $ H_{1} $ , $ V_{1}$ represent the horizontal and vertical components of $ mvp_{1} $, and $ abs(x) $ is the absolute value of $ x $. The $ T $ will be set at 0, 1, 5, 20, and 1000 for experiments. The embedding capacity for Tar3 is bpap (Bits Per AMVP PU), which shall be set at 0.1, 0.2, 0.3, 0.4, and 0.5. All the steganography methods are implemented based on the official test model HM16.9\cite{ref28}. 

\subsubsection{Competitor Steganalysis Methods}
There are two types of competitor steganalysis methods used for the experiments. 
The first type is the parallel porting of classical methods proposed for H.264/AVC to HEVC, including AoSO \cite{ref19} and NPELO \cite{ref20}. 
The other type is the MV-based steganalysis methods proposed based on HEVC, including the neural network-based VSRNet method \cite{ref23} and the local optimality in the candidate list-based LOCL method \cite{ref26}. 
All the steganalysis feature sets are extracted based on the official test model HM16.9.

\subsubsection{Training and Classification}
It is worth noting that the steganalysis scheme proposed in this article does not require the use of machine learning methods for training and classification, as we can determine whether there exist message embedding based on the proposed optimal rate of MVP. 
To implement training and classification for various competitive steganalysis approaches, we use a Gaussian-kernel SVM (support vector machine)\cite{ref29}, whose penalty factor C and kernel factor are established via a five-fold cross-validation. Additionally, the accuracy rate—which is calculated as the proportion of correctly identified samples to all samples—is used to gauge the effectiveness of the detection process. Ten randomly selected database splits are used to get the final accuracy rate. Each iteration uses 40\% of the cover-stego video pairings for testing, and 60\% are randomly selected for training. A desktop computer with a 3.1GHz Intel Core i9 CPU and 64 GB RAM is used to conduct all of the tests. 

\subsection{The Optimal Rate of MVP for Cover video}
This experiment verifies the applicability of Property 1 of Section \ref{chap:Proposed method} under different conditions, i.e., the optimal rate of MVP is calculated according to Formula (\ref{formula7}) for cover video in DB1 and DB2 databases. To detect the impact of different encoders, the cover videos in this experiment are compressed using two different encoders.

The first encoder is HM16.9, and the cover videos are encoded with quality parameters (QP) of 20, 25, and 30. The GOP (Group Of Picture) structure used for HM is IPPPPPPPPPI..., and the experimental results are shown in TABLE \ref{tbl_cover_HM}. Two metrics are counted in the table, the first is the average of the optimal rate of MVP of all videos in the database, and the second is the proportion of videos with a 100\% optimal rate to the whole videos. The experimental data shows that both DB1 and DB2, $Optimal(mvp)$ have a mean value of 100\% at different QPs. The proportion of videos with a 100\% optimal rate to the whole videos is also 100\%. The experimental data indicate that among the cover videos encoded with HM, all the PU encoded with AMVP meet the local optimality of MVP, i.e., they satisfy the Property 1.

\begin{table}[h]
	\caption{\label{tbl_cover_HM} The optimal rate of MVP for cover videos encoded by HM.}
	\centering
	\renewcommand\arraystretch{1.3}
	\begin{tabular}{cccc}
		\toprule
		Database  & QP  &$Optimal(mvp)$  &  \makecell{proportion of videos with\\ 100\% optimal rate}\\
		\toprule 		
		\multirow{3}{*}{DB1}	                  &20	        &100	  &100     \\
		   &25	        &100	  &100     \\
		  &30	        &100	  &100     \\
		\midrule
		\multirow{3}{*}{DB2}	                  &20	        &100	  &100     \\
		   &25	        &100	  &100     \\
		   &30	        &100	  &100     \\
		\bottomrule                 	
	\end{tabular} 
\end{table}

The second encoder is the efficient x265\cite{ref30}. The parameters used for x265 are the same as HM above. The experimental results are shown in Table \ref{tbl_cover_x265}, and it can be seen that both in DB1 and DB2, $Optimal(mvp)$ have a mean value of 100\% at different QPs. The proportion of videos with a 100\% optimal rate to the whole videos is also 100\%. The experimental result indicates that the cover video compressed by x265 also satisfies the Property 1.
\begin{table}[h]
	\caption{\label{tbl_cover_x265} The optimal rate of MVP for cover videos encoded by x265.}
	\centering
	\renewcommand\arraystretch{1.3}
	\begin{tabular}{cccc}
		\toprule
		Database  & QP  &$Optimal(mvp)$  &  \makecell{proportion of videos with\\ 100\% optimal rate}\\
		\toprule 		
		\multirow{3}{*}{DB1}	                  &20	        &100	  &100     \\
		&25	        &100	  &100     \\
		&30	        &100	  &100     \\
		\midrule
		\multirow{3}{*}{DB2}	                  &20	        &100	  &100     \\
		&25	        &100	  &100     \\
		&30	        &100	  &100     \\
		\bottomrule                 	
	\end{tabular} 
\end{table}

The results of the above two experiments show that both the official reference software HM and the optimized high-performance encoder x265 have an optimal rate of 100\% of MVP for cover videos under different encoding parameters, which follows the Property 1.

\subsection{The Optimal Rate of MVP for Stego video}\label{chap:Optimal rate of stego}
This experiment analyzes the detection performance of the proposed method on stego video. We use different steganography methods for message embedding on DB1 and DB2 databases, with an encoder of HM16.9 and a GOP structure of IPPPPPPPI...

The experimental results for the steganography algorithm Tar1\cite{ref14} are shown in TABLE \ref{tbl_stego_tar1}. Taking the embedding strength $ e=0.1 $ at QP=20 in the DB1 database as an example, the average value of $Optimal(mvp)$ is 99.70\%, which means the optimality of the MVP of 0.3\% PUs is corrupted. The proportion of videos with $Optimal(mvp)$ equal to 100\% of the total videos is 0\%, which indicates that no video fully satisfies the MVP optimality. 
As a whole from TABLE \ref{tbl_stego_tar1}, the average value of $Optimal(mvp)$ is lower than 100\% for different databases and QPs, and the proportion of the number of videos with $Optimal(mvp)$ equal to 100\% is 0\%, which indicates that the optimality of MVP of all videos is perturbed. 
Based on the above analysis, it can be determined that all video sequences in this experiment are stego videos, and the reason is that the embedding operation performed by Tar1 on the MVDs destroys the local optimality of the MVP, which is consistent with the theoretical analysis in Section \ref{chap:MV-Steganography-mvd} and Property 2 in Section \ref{chap:Proposed method}.
\begin{table}[h]
	\caption{\label{tbl_stego_tar1} Statistical data of the optimal rate of MVP for steganography method Tar1 \cite{ref14}. The first value is the average (\%) of the optimal rate of all videos, and the second value is the proportion (\%) of the videos with a 100\% optimal rate to the whole videos.}
	\centering
	\renewcommand\arraystretch{1.3}
	\begin{tabular}{ccccccc}
		\toprule
		\multirow{2}{*}{Database}  & \multirow{2}{*}{QP}   &\multicolumn{5}{c}{Embedding Strength $ e $ }    \\
		  \cline{3-7}   
		                           &                       &0.1        &0.2        &0.3        &0.4        &0.5\\      
		                              
		\toprule 		
		\multirow{3}{*}{DB1}	   &20	                   &99.70	 0	&99.60	 0	&99.35	 0	&99.29	 0	&99.14	 0 \\
		                           &25	                   &99.85	 0	&99.70	 0	&99.52	 0	&99.41	 0	&99.24	 0 \\
	                           	   &30	                   &99.87	 0	&99.73	 0	&99.58	 0	&99.47	 0	&99.33	 0\\
		\midrule
		\multirow{3}{*}{DB2}	   &20	                   &99.81	 0	&99.70	 0	&99.57	 0	&99.47	 0	&99.38	 0 \\
									&25	                   &99.83	 0	&99.70	 0	&99.54	 0	&99.40	 0	&99.28	 0 \\
									&30	                   &99.85	 0	&99.72	 0	&99.59	 0	&99.45	 0	&99.34	 0\\
		\bottomrule                 	
	\end{tabular} 
\end{table}

The experimental results for the steganography algorithm Tar2\cite{ref16} are shown in TABLE \ref{tbl_stego_tar2}. 
With different databases and QPs, the average value of $Optimal(mvp)$ is 100\% when the embedding threshold $ T=0 $ and the proportion of videos with $Optimal(mvp)$ equal to 100\% of the total videos is 0\%. According to the definition of $ T $ in Section \ref{chap:steganography methods}, $ T=0 $ means that $ mvp_{0} $ is the same as $ mvp_{1} $, so modifying the MVP index does not change the optimality of the MVP. That means when $ T=0 $, the scheme of this paper is invalid. In fact, at $ T=0 $, the Tar2 algorithm only selects those PUs whose $ mvp_{0} $ is the same as $ mvp_{1} $ for message embedding, and its embedding capacity is smaller. 
When $ T\neq0 $, the first indicator in the experimental results are not 100\%, and the second indicator are 0\%, which indicates that the optimality of the MVP of all videos is perturbed. 
We can determine these videos as stego based on these two statistical indicators. The reason for the above experimental results is that the embedding operation performed by Tar2 on the MVP index destroys the MVP's local optimality, which is consistent with the theoretical analysis in Section \ref{chap:MV-Steganography-idx} and Property 2 in Section \ref{chap:Proposed method}.

\begin{table}[h]
	\caption{\label{tbl_stego_tar2} Statistical data of the optimal rate of MVP for steganography method Tar2 \cite{ref16}. The first value is the average (\%) of the optimal rate of all videos, and the second value is the proportion (\%) of the videos with a 100\% optimal rate to the whole videos.}
	\centering
	\renewcommand\arraystretch{1.3}
	\begin{tabular}{ccccccc}
		\toprule
		\multirow{2}{*}{Database}  & \multirow{2}{*}{QP}   &\multicolumn{5}{c}{Embedding Threshold $ T $ }    \\
		\cline{3-7}   
		&                       &0        &1        &5        &20        &1000\\      
		
		\toprule 		
		\multirow{3}{*}{DB1}	   &20	                   &100	 100	&99.90	 0	&82.27	 0	&72.84	 0	&69.17	 0 \\
									&25	                   &100	 100	&93.27	 0	&81.92	 0	&71.30	 0	&66.80	 0 \\
									&30	                   &100	 100	&94.04   0	&82.97	 0	&71.55	 0	&66.36	 0\\
		\midrule
		\multirow{3}{*}{DB2}	   &20	                   &100	 100	&95.17	 0	&84.85	 0	&75.01	 0	&68.57	 0 \\
									&25	                   &100	 100	&94.65	 0	&83.34	 0	&72.31	 0	&65.77	 0 \\
									&30	                   &100	 100	&95.02	 0	&83.69	 0	&71.52	 0	&64.07	 0\\
		\bottomrule                 	
	\end{tabular} 
\end{table}

\begin{table}[b]
	\caption{\label{tbl_stego_tar3} Statistical data of the optimal rate of MVP for steganography method Tar3 \cite{ref17}. The first value is the average (\%) of the optimal rate of all videos, and the second value is the proportion (\%) of the videos with a 100\% optimal rate to the whole videos.}
	\centering
	\renewcommand\arraystretch{1.3}
	\begin{tabular}{ccccccc}
		\toprule
		\multirow{2}{*}{Database}  & \multirow{2}{*}{QP}   &\multicolumn{5}{c}{Embedding Capacity(in bpap)}    \\
		\cline{3-7}   
		&                       &0.1        &0.2        &0.3        &0.4        &0.5\\      
		\toprule 		
		\multirow{3}{*}{DB1}	   &20	                   &98.88	0	&98.99	 0	&98.08	 0	&96.46	 0	&94.40	 0 \\
		&25	                   &98.17	0	&98.49	 0	&97.69	 0	&95.98	 0	&93.74	 0 \\
		&30	                   &94.46	0	&96.24	 0	&95.53	 0	&94.17	 0	&92.03	 0\\
		\midrule
		\multirow{3}{*}{DB2}	   &20	                   &99.46	0	&99.26	 0	&98.38	 0	&96.78	 0	&94.59  0 \\
		&25	                   &99.49	0	&99.17	 0	&98.06	 0	&96.15	 0	&93.64	 0 \\
		&30	                   &99.08	0	&98.91	 0	&97.56	 0	&95.32	 0	&92.56	 0\\
		\bottomrule                 	
	\end{tabular} 
\end{table}

The experimental results for the steganography algorithm Tar3 \cite{ref17} are shown in TABLE \ref{tbl_stego_tar3}. The experimental results are similar to those of Tar2, but the Tar3 method is an adaptive embedding in MVP index with STC. The experimental results show that the proposed scheme can ideally detect the damage to the local optimality of the MVP. The underlying reason is that the embedding operation performed by Tar3 on the MVP index destroys the local optimality of the MVP, which is also consistent with the theoretical analysis in Section \ref{chap:MV-Steganography-idx} and Property 2 in Section \ref{chap:Proposed method}.

Through the above analysis, for the three state-of-the-art steganography methods, the proposed steganalysis feature in this paper is invalid only in the case of $ T=0 $ for the Tar2. In all other conditions, we can accurately distinguish the cover video from the stego video by whether the optimal rate of MVP is equal to 100\%. The above experimental findings also verify the correctness of Property 2 and Corollary 1 in Section \ref{chap:Proposed method}.

\begin{table*}[t]
	\caption{\label{tbl_stego_competitor} Detector accuracy(\%) of competitor steganalysis methods against 3 steganography methods in database DB1.}
	\centering
	\renewcommand\arraystretch{1.3}
	\begin{tabular}{ccccccccc}
		\toprule
		\multirow{2}{*}{\makecell{Steganalysis\\ Methods}}  & \multirow{2}{*}{QP}   
		&\multirow{2}{*}{\makecell{Steganography Method\& \\Embedding Parameter} }  
		&\multicolumn{2}{c}{\makecell{Tar1\cite{ref14} \\Embedding Strength $ e $}}  
		&\multicolumn{2}{c}{\makecell{Tar2\cite{ref16} \\Embedding Threshold $ T $}}
		&\multicolumn{2}{c}{\makecell{Tar3\cite{ref17} \\Embedding Capacity(in bpap)}}  \\
		\cline{4-9} 
		&      &       &0.1      &0.5      &0       &1000      &0.1      &0.5  \\
		\toprule 		
		
		\multirow{2}{*}{AoSO\cite{ref19}}		 &20	&       & 58.29	  &67.80	&49.17	 &48.17	    &48.45	  &50.04 \\
		&30	&       & 54.92	&63.90	&49.20	&48.80	&48.90	&48.15 \\
		\midrule									 
		\multirow{2}{*}{NPELO\cite{ref20}}	 &20	&       & 71.93	&74.39	&63.47	&73.89	&63.22	&60.99 \\
		&30	&       & 68.31	&70.44	&60.86	&71.64	&45.40	&48.88 \\
		\midrule
		\multirow{2}{*}{VSRNet\cite{ref23}}	 &20	&       & 68.11	&70.19	&49.34	&56.12	&49.34	&52.42 \\
		&30	&       & 65.24	&67.83	&50.23	&54.90	&49.08	&51.98\\
		\midrule
		\multirow{2}{*}{LOCL\cite{ref26}}	 &20	&       & 72.36	&76.80	&59.96	&72.82	&69.55	&68.21 \\
		&30	&       & 65.77	&67.56	&46.21	&87.32	&57.74	&64.62 \\
		\bottomrule                 	
	\end{tabular} 
\end{table*}

\subsection{Comparison with other Machine Learning-based Steganalysis Methods}
To compare the detection performance of existing steganalysis methods against the above three steganography algorithms, this section uses the existing state-of-the-art steganalysis methods to detect the stego video in Section \ref{chap:Optimal rate of stego}. Due to limitations in the length of the paper, we only list some experimental data on database DB1 in Table \ref{tbl_stego_competitor}. 
The feature set of AoSO\cite{ref19} uses the SAD to describe the MV's local optimal. From the experimental data, AoSO has some detection effect on Tar1 because Tar1 is a steganography method that directly modifies the MV and destroys the local optimality of the MV. AoSO is ineffective on Tar2 and Tar3 because these two steganography methods embed messages in the MVP index, and the original MV remains unchanged. 
NPELO\cite{ref20} is a steganalysis feature set based on the local optimality of MV rate-distortion. The experimental data shows that NPELO performs better in detecting Tar1 than Tar2 and Tar3 for reasons similar to AoSO. The overall performance of NPELO is better than that of AoSO because NPELO considers rate-distortion (including pixel distortion and code bit), making it more reasonable. 
VSRNet\cite{ref23} is a neural network-based steganalysis method, which is effective in detecting Tar1 but ineffective for Tar2 and Tar3. The experimental results indicate that VSRNet cannot yet capture the perturbations caused by the steganography method, which embeds messages in the index of MVP.
LOCL\cite{ref26} is a feature set that combines the NPELO and the optimal MVP candidate list, and its performance is better than AoSO, NPELO and VSRNet overall. However, LOCL considers the MVP's optimality together with the MV's optimality when designing the features, but the optimality of the MVP still needs to be fully exploited.

From the above analysis, the detection accuracy of traditional steganalysis features for the three steganography algorithms is low (mostly below 80\%). In contrast, according to the analysis in Section \ref{chap:Optimal rate of stego}, the proposed optimal rate of MVP can perfectly distinguish cover vector video from stego video in most cases. And more importantly, the proposed method in this paper does not need to train the classifier as the experiments in this section, so it is more practical and efficient.

\subsection{Applicability in B-Frames}
The proposed method is based on the AMVP technique and can be applied to all videos encoded with the AMVP technique. Therefore, although the inter-Frame coding frames used in the previous experiments are P-Frames, they are also theoretically applicable to B-Frames. If there are two reference lists for the PU block encoded by the AMVP technique in the B-Frame, the MVP on both reference lists satisfies the properties of Section \ref{chap:Proposed method}.

To verify the applicability of the proposed scheme on B-Frames, we use the GOP structure of IBBBBBBBBI... for this experiment. The experiment is performed on the DB1 database with the steganography method as Tar2 (it is worth noting that Tar1 and Tar3 are also applicable and are not listed in this paper due to space limitation), and other parameters are the same as in Section \ref{chap:Optimal rate of stego}. TABLE \ref{tbl_stego_B_Frames} shows the statistical results of the optimal rate of MVP for the cover video and stego video. 
As can be seen from the data in the table, the average value of $Optimal(mvp)$ for the cover videos at different QPs is 100\%, and the videos with 100\% optimal rate account for 100\% of the whole dataset, indicating that the experimental results at B-Frame also satisfy the Property 1 in Section \ref{chap:Proposed method}. 

The results at the embedding threshold $ T=0 $ are consistent with those at $ T=0 $ in Table \ref{tbl_stego_tar2}, again because only the PUs whose $ mvp_{0}=mvp_{1} $ are used for message embedding, so the optimality of the MVP is not destroyed. In contrast, when $T\neq0$, $Optimal(mvp)$ in the experimental results is not 100\%, and the proportion of the videos with a 100\% optimal rate to the whole videos is 0\%, which indicates that the optimality of the MVP of all videos is perturbed. In summary, for the HEVC videos compressed with B-Frames, except for $ T=0 $, the proposed optimal rate of MVP can still effectively distinguish between cover and stego videos.

\begin{table}[t]
	\caption{\label{tbl_stego_B_Frames} Statistical results for the cover and stego video with GOP structure of IBBBBBBBBI.... The steganography algorithm is Tar2 \cite{ref16}, and the database is DB1. The first value is the average (\%) of the optimal rate of all videos, and the second value is the proportion (\%) of the videos with a 100\% optimal rate to the whole videos.}
	\centering
	\renewcommand\arraystretch{1.3}
	\begin{tabular}{ccccccc}
		\toprule
		\multirow{2}{*}{QP}  & \multirow{2}{*}{Cover}   &\multicolumn{5}{c}{Embedding threshold $ T $}    \\
		\cline{3-7}   
		&                       &0        &1        &5        &20        &1000\\      
		\toprule 		
									20	    &100  100	&100  100	&95.80	 0	&87.44	 0	&79.63	 0	&75.97	 0 \\
									25	    &100  100	&100  100	&96.16	 0	&87.52	 0	&77.94	 0	&72.45	 0 \\
									30	    &100  100	&100  100	&96.70	 0	&88.05	 0	&77.59	 0	&71.28	 0\\
		\bottomrule                 	
	\end{tabular} 
\end{table}

\subsection{The Complexity Analysis of the Proposed Feature}
In order to analyze the computational complexity of the proposed scheme, this subsection compares the time required for feature extraction with different QPs. Table 9 shows the dimensionality of the four steganalysis features and the average time needed to extract a video sequence (see Section \ref{chap:Optimal rate of stego} for parameter settings, CIF format, 240 frames). The experiments are run on a desktop computer with a 3.1GHz Intel Core i9 CPU and 64 GB RAM. 
The data in the table shows that the feature dimension of the proposed scheme in this paper is only 1, which is the lowest among all methods. Regarding computational complexity, both AoSO and NPELO need to compute the 1-neighbourhood optimality of the MV, and the computational complexity is close. The highest complexity is LOCL because it has to calculate not only the optimality of the MV but also the optimality of the MVP’s candidate list. The extraction time of the proposed scheme is only about 1/2 of the other algorithms, because the proposed scheme does not need to calculate the 1-neighborhood optimality of the MV, but only the rate-distortion of the two MVPs. In addition, the smaller the QP, the larger the running time of all algorithms. This is because the smaller the QP, the finer the division of coding blocks, the more MVs in the code stream, and the more data to be processed.

Overall, the computational complexity of the scheme in this paper is minimal because the feature dimension is only one, and the rate-distortion of only two MVPs needs to be calculated. More importantly, the proposed scheme does not require extensive machine learning based training. Therefore, the proposed method is very efficient and can be applied to practical scenarios.

\begin{table}[h]
	\caption{\label{tbl_stego_complexity} Average computational time (in second) of feature extraction for a single CIF sequence with 240 frames.}
	\centering
	\renewcommand\arraystretch{1.3}
	\begin{tabular}{ccccc}
		\toprule
		\multirow{2}{*}{Steganalysis Methods}  & \multirow{2}{*}{Dimension}   &\multicolumn{3}{c}{QP}    \\
		\cline{3-5}   
		&                       &20        &25        &30        \\      
		\toprule 		
		AoSO\cite{ref19}	    &18	&4.15	&3.30	&2.64	 \\
		NPELO\cite{ref20}	    &36	&4.45	&3.55	&2.86	 \\
		LOCL\cite{ref26}	    &37	&4.70	&3.85	&2.95	\\
		Proposed				&1	&2.40	&1.81	&1.39   \\
		\bottomrule                 	
	\end{tabular} 
\end{table}

\section{Conclusion}
The development of video coding standards always aims to reduce redundancy and increase compression performance while ensuring visual quality\cite{ref13,ref_VVC}. In contrast, steganography's fundamental starting point is exploiting data redundancy to embed messages. Therefore although new coding standards provide more coding elements (e.g., MVP index in AMVP techniques) for message embedding, the reduction of redundancy also poses challenges to steganography. For example, with the widespread adoption of HEVC video standard, video steganography and steganalysis based on HEVC have received more and more attention. Although the AMVP technique provides more embedding space for steganography, it also exposes risks. 

We analyze that either the MVP index or the MVD for message embedding may lead to perturbations in the optimality of the MVP. Based on the above observation, we design an optimal rate of MVP with dimension only one as a steganalysis feature. This feature can accurately distinguish cover videos from stego videos and has the advantage of low complexity. In the subsequent work, we will focus on taking advantage of the AMVP technique to embed messages in HEVC while ensuring that the optimality of the MVP is not destroyed, and achieving the goal of improving the embedding capacity while enhancing the security against steganalysis.

\section{Acknowledgements} 
This work is supported by the National Natural Science Foundation of China (Grant No.62272478, No.62202496, No.62102450).

\newpage

\vfill


\begin{thebibliography}{1}
\bibliographystyle{IEEEtran}

\bibitem{ref1}
J. Li, M. Zhang, K. Niu, and X. Yang, ‘A Review of Motion Vector-Based Video Steganography’, Secur. Commun. Networks, vol. 2022, pp. 1–19, 2022, doi: 10.1155/2022/2946812.

\bibitem{ref2}
Y. Dong, X. Jiang, Z. Li, T. Sun, and Z. Zhang, ‘Multi-Channel HEVC Steganography by Minimizing IPM Steganographic Distortions’, IEEE Trans. Multimed., vol. 25, pp. 2698–2709, 2023.

\bibitem{ref3}
Y. Dong, X. Jiang, Z. Li, T. Sun, and P. He, ‘Adaptive HEVC Steganography Based on Steganographic Compression Efficiency Degradation Model’, IEEE Trans. Dependable Secur. Comput., vol. 20, no. 1, pp. 769–783, 2023.

\bibitem{ref4}
H. Zhang, Y. Cao, X. Zhao, W. Zhang, and N. Yu, ‘Video steganography with perturbed macroblock partition’, in Proceedings of the 2nd ACM workshop on Information hiding and multimedia security - IH\&MMSec ’14, 2014, pp. 115–122.

\bibitem{ref5}
Z. Li, X. Jiang, Y. Dong, L. Meng, and T. Sun, ‘An Anti-Steganalysis HEVC Video Steganography With High Performance Based on CNN and PU Partition Modes’, IEEE Trans. Dependable Secur. Comput., vol. 20, no. 1, pp. 606–619, 2023.

\bibitem{ref6}
J. Liu, Z. Li, X. Jiang, and Z. Zhang, ‘A High-Performance CNN-Applied HEVC Steganography Based on Diamond-Coded PU Partition Modes’, IEEE Trans. Multimed., vol. 24, pp. 2084–2097, 2022.

\bibitem{ref7}
J. Wang, X. Jia, X. Kang, and Y. Q. Shi, ‘A Cover Selection HEVC Video Steganography Based on Intra Prediction Mode’, IEEE Access, vol. 7, pp. 119393–119402, 2019.

\bibitem{ref8}
S. He, D. Xu, L. Yang, and Y. Liu, ‘HEVC video information hiding scheme based on adaptive double-layer embedding strategy’, J. Vis. Commun. Image Represent., 2022, doi: 10.1016/j.jvcir.2022.103549.

\bibitem{ref9}
J. Li, M. Zhang, K. Niu, and X. Yang, ‘Investigation on principles for cost assignment in motion vector-based video steganography’, J. Inf. Secur. Appl., vol. 73, p. 103439, 2023, doi: 10.1016/j.jisa.2023.103439.

\bibitem{ref10}
M. Guo, T. Sun, X. Jiang, Y. Dong, and K. Xu, ‘A Motion Vector-Based Steganographic Algorithm for HEVC with MTB Mapping Strategy’, in International Workshop on Digital Watermarking 2019, 2019, pp. 293–306.

\bibitem{ref_Liu_tcsvt_2022}
Y. Liu, J. Ni, W. Zhang, and J. Huang, ‘A Novel Video Steganographic Scheme Incorporating the Consistency Degree of Motion Vectors’, IEEE Trans. Circuits Syst. Video Technol., vol. 32, no. 7, pp. 4905–4910, 2022.

\bibitem{ref11}
Y. Chen et al., ‘DDCA: A Distortion Drift-Based Cost Assignment Method for Adaptive Video Steganography in the Transform Domain’, IEEE Trans. Dependable Secur. Comput., vol. 19, no. 4, pp. 2405–2420, 2022.

\bibitem{ref12}
Y. Wang, Y. Cao, and X. Zhao, ‘CEC: Cluster Embedding Coding for H.264 Steganography’, IEEE Signal Process. Lett., vol. 27, pp. 955–959, 2020.

\bibitem{ref13}
G. J. Sullivan, J.-R. Ohm, W.-J. Han, and T. Wiegand, “Overview of the High Efficiency Video Coding (HEVC) Standard,” IEEE Transactions on Circuits and Systems for Video Technology, vol. 22, no. 12, pp. 1649–1668, 2012.

\bibitem{ref14}
J. Yang and S. Li, ‘An efficient information hiding method based on motion vector space encoding for HEVC’, Multimed. Tools Appl., vol. 77, no. 10, pp. 11979–12001, 2018.

\bibitem{ref16}
Y. Hu, W. Gong, F. Liu, L. Liu, and M. Zhu, ‘Large-Capacity Lossless HEVC Information Hiding Based on Index Parameter Modification’, J. South China Univ. Technol. ( Nat. Sci. Ed.), vol. 46, no. 5, pp. 1–8, 2018.

\bibitem{ref17}
S. Liu, B. Liu, Y. Hu, and X. Zhao, ‘Non-Degraded Adaptive HEVC Steganography by Advanced Motion Vector Prediction’, IEEE Signal Process. Lett., vol. 28, pp. 1843–1847, 2021.

\bibitem{ref18}
T. Filler, J. Judas, and J. Fridrich, ‘Minimizing additive distortion in steganography using syndrome-trellis codes’, IEEE Trans. Inf. Forensics Secur., vol. 6, no. 3 PART 2, pp. 920–935, 2011.

\bibitem{ref19}
K. Wang, H. Zhao, and H. Wang, ‘Video steganalysis against motion vector-based steganography by adding or subtracting one motion vector value’, IEEE Trans. Inf. Forensics Secur., vol. 9, no. 5, pp. 741–751, 2014.

\bibitem{ref20}
H. Zhang, Y. Cao, and X. Zhao, ‘A steganalytic approach to detect motion vector modification using near-perfect estimation for local optimality’, IEEE Trans. Inf. Forensics Secur., vol. 12, no. 2, pp. 465–478, 2017.

\bibitem{ref21}
L. Zhai, L. Wang, and Y. Ren, ‘Universal Detection of Video Steganography in Multiple Domains Based on the Consistency of Motion Vectors’, IEEE Trans. Inf. Forensics Secur., vol. 15, no. c, pp. 1762–1777, 2020.

\bibitem{ref22}
T. Shanableh, ‘Feature extraction and machine learning solutions for detecting motion vector data embedding in HEVC videos’, Multimed. Tools Appl., vol. 80, no. 18, pp. 27047–27066, 2021.

\bibitem{ref23}
X. Huang, Y. Hu, Y. Wang, B. Liu, and S. Liu, ‘Deep Learning-based Quantitative Steganalysis to Detect Motion Vector Embedding of HEVC Videos’, in 2020 IEEE Fifth International Conference on Data Science in Cyberspace (DSC), Jul. 2020, no. 2019, pp. 150–155.

\bibitem{ref24}
X. Huang, Y. Hu, Y. Wang, B. Liu, and S. Liu, ‘Selection-Channel-Aware Deep Neural Network to Detect Motion Vector Embedding of HEVC Videos’, ICSPCC 2020 - IEEE Int. Conf. Signal Process. Commun. Comput. Proc., 2020, doi: 10.1109/ICSPCC50002.2020.9259551.

\bibitem{ref25}
Y. Hu, X. Huang, Y. Wang, B. Liu, and S. Liu, ‘Improving deep learning-based video steganalysis by using MVD detection space’, Journal of image and graphics., vol. 03, no. 28, pp. 702-715, 2022.

\bibitem{ref26}
S. Liu, Y. Hu, B. Liu, and C.-T. Li, ‘An HEVC steganalytic approach against motion vector modification using local optimality in candidate list’, Pattern Recognit. Lett., vol. 146, pp. 23–30, 2021.

\bibitem{ref27}
YUV Video Sequences. Accessed: Feb. 2022. [Online]. Available: http://trace.eas.asu.edu/yuv/index.html

\bibitem{ref28}
HEVC HM Reference Software. Accessed: Feb. 2022. [Online]. Available: http://www.hevc.info

\bibitem{ref29}
C.-C. Chang and C.-J. Lin, LIBSVM: a library for support vector machines. [Online]. Available: http://www.csie.ntu.edu. tw/$ \sim $cjlin/libsvm

\bibitem{ref30}
VideoLAN, x265. Accessed: Feb. 2022. [Online]. Available: https://www.videolan.org/developers/x265.html

\bibitem{ref_VVC}
B. Bross et al., ‘Overview of the Versatile Video Coding (VVC) Standard and its Applications’, IEEE Trans. Circuits Syst. Video Technol., vol. 31, no. 10, pp. 3736–3764, 2021.
\end{thebibliography}
\end{document}